\documentclass[longauth]{aa}     

\usepackage{graphicx}
\usepackage{txfonts}
\usepackage{subcaption}         
\usepackage{lscape}             
\usepackage{placeins}           
                                
\usepackage{hyperref}
\hypersetup{
    colorlinks=true,
    linkcolor=blue,
    filecolor=blue,
    urlcolor=blue,
    pdftitle={J1447-0149},
    citecolor=blue,
    } 
\usepackage[dvipsnames]{xcolor}

\AtBeginDocument{%
  \let\runningpagewiselinenumbers\relax
  \let\runninglinenumbers\relax
  
  \let\linenumbers\relax
  \nolinenumbers
}

\begin{document}

   \title{KiDS J1447-0149: The first spatially resolved spectroscopy of a relic galaxy beyond the local universe }

   \subtitle{An old high-dispersion core embedded in a compact rotating stellar structure}


   \author{Johanna Hartke\inst{\ref{FINCA},\ref{UTU}, \ref{TCSMT}}\corrauth{johanna.hartke@utu.fi}
        \and 
        Chiara Spiniello\inst{\ref{ESO-G}, \ref{Oxford}, \ref{INAF-OAC}}\email{chiara.spiniello@eso.org} 
        \and 
        Michele Cappellari\inst{\ref{Oxford}}\email{michele.cappellari@physics.ox.ac.uk}
        \and 
        Davide Bevacqua\inst{\ref{INAF-Rome}}\email{davide.bevacqua@inaf.it}
        \and 
        Enrico Congiu\inst{\ref{ESO-CL}}\email{econgiu@eso.org}
        \and
        Anna Ferré-Mateu\inst{\ref{IAC},\ref{ULL}}\email{aferremateu@gmail.com} 
        \and
        Adriano Poci\inst{\ref{Oxford}}\email{adriano.poci@physics.ox.ac.uk}
        \and 
        Claudia Pulsoni\inst{\ref{MPE}}\email{cpulsoni@mpe.mpg.de}
        \and 
        Magda Arnaboldi\inst{\ref{ESO-G}}\email{marnabol@eso.org}
        \and
        Giuseppe D'Ago\inst{\ref{Cambridge}}\email{gdago.astro@gmail.com}
        \and
        Michalina Maksymowicz-Maciata\inst{\ref{Bristol}}\email{michalina.maksymowicz.maciata@gmail.com}
        \and
        Paolo Saracco\inst{\ref{INAF-Milano}}\email{paolo.saracco@inaf.it}
        \and 
        Diana Scognamiglio\inst{\ref{Duke}}\email{diana.scognamiglio@duke.edu}
        \and
        David A. Simon\inst{\ref{Westlake}}\email{davidsimon@westlake.edu.cn}
        \and 
        Crescenzo Tortora\inst{\ref{INAF-OAC}}\email{crescenzo.tortora@inaf.it}
        \and
        Petri Väisänen\inst{\ref{FINCA}, \ref{SAAO}}\email{petri.vaisanen@utu.fi}}

   \institute{Finnish Centre for Astronomy with ESO,
                (FINCA), University of Turku, 20014 Turku, Finland \label{FINCA}             
                \and
                Tuorla Observatory, Department of Physics and Astronomy, University of Turku, 20014 Turku, Finland \label{UTU} \and 
                Turku Collegium for Science, Medicine and Technology (TCSMT), University of Turku, FI-20014 Turku, Finland \label{TCSMT}
                \and
                European Southern Observatory, Karl-Schwarzschild-Straße 2, 85748 Garching bei München, Germany \label{ESO-G} 
                \and 
                Sub-Department of Astrophysics, Department of Physics, University of Oxford, Keble Road, Oxford OX1 3RH, United Kingdom \label{Oxford}
                \and 
                INAF -- Osservatorio Astronomico di Capodimonte, Via Moiariello 16, 80131 Naples, Italy \label{INAF-OAC}
                \and 
                INAF -- Osservatorio Astronomico di Roma, Via Frascati 33, 00078 Monte Porzio Catone, Italy \label{INAF-Rome}
                \and 
                European Southern Observatory, Alonso de Córdova 3107, Casilla 19, Santiago 19001, Chile \label{ESO-CL}
                \and
                Instituto de Astrof\'isica de Canarias, Calle V\'ia L\'actea s/n, 38205. La Laguna, Tenerife, Spain \label{IAC}
                \and
                Departamento de Astrof\'isica, Universidad de La Laguna, E-38200, La Laguna, Tenerife, Spain \label{ULL}
               \and 
                Max-Planck-Institut f\"{u}r Extraterrestrische Physik, 
                Giessenbachstra{\ss}e, 85748 Garching, Germany\label{MPE} 
                \and
                Institute of Astronomy, University of Cambridge, Madingley Road, Cambridge CB3 0HA, United Kingdom \label{Cambridge}
                \and 
                School of Physics, H.H. Wills Physics Laboratory, Tyndall Avenue, University of Bristol, Bristol BS8 1TL, United Kingdom \label{Bristol}
                \and
                INAF -- Osservatorio Astronomico di Brera, via Brera 28, 20121 Milano, Italy \label{INAF-Milano}
                \and
                Duke University, Durham, NC 27708, USA\label{Duke}
                \and
                Department of Astronomy, Westlake University, Hangzhou 310030, Zhejiang Province, People’s Republic of China \label{Westlake}
                \and
                South African Astronomical Observatory, PO Box 9, Observatory 7935, South Africa \label{SAAO}
             }

   \date{\today}

 
\abstract
{Massive relic galaxies are considered the surviving descendants of high-redshift compact quiescent systems. Integrated spectroscopy has identified relic candidates beyond the local Universe, but their internal kinematic and stellar-population structure remains largely unexplored.}
{We present the first spatially resolved spectroscopic study of KiDS J1447-0149, a massive relic galaxy  at $z=0.2074$, observed with MUSE and adaptive optics. 
 We aim to characterize its internal kinematics and stellar population properties with the goal of probing its spatially resolved mass assembly and star formation history.}
{We measured spatially resolved stellar kinematics with full-spectral fitting from binned data, with sub-kpc bin sizes, covering $\sim 1.4$ effective radii. 
We then used a coarser three-bin configuration informed by the kinematics to recover the higher-order moments of the velocity distribution and to derive line-strength-based $[\alpha/{\rm Fe}]$, as well as stellar age and metallicity from full-spectral fitting. The latter were used to reconstruct the star formation history (SFH). We also computed the degree of relicness (DoR), a dimensionless quantity that measures how early and rapidly each system formed its stellar mass, from the SFH.} 
{The MUSE-NFM data reveal a coherent velocity gradient, showing that J1447-0149 is not purely pressure supported. The velocity-dispersion field displays a central enhancement, reaching $\sigma_\star=233\pm13\,{\rm km\,s^{-1}}$, substantially larger than previous seeing-limited measurements ($\sigma_{\star, \rm INSPIRE} =187 \pm 9\,{\rm km\,s^{-1}}$). The $h_3$--$V_\star$ anti-correlation and mildly positive $h_4$ values support a composite structure, with a rotating stellar component surrounding a compact dynamically hot core. The resolved stellar populations show that this central, dispersion-dominated region is also the oldest and most metal-rich component. Its SFH rises rapidly, with the stellar mass assembled within $\sim2\,{\rm Gyr}$ after the Big Bang, and reaches the highest DoR, ${\rm DoR}=0.9^{+0.1}_{-0.2}$. The two outer bins have a DoR value of ${\rm DoR}=0.8\pm0.2$, consistent with the central bin, but possibly indicating slightly longer formation times.}
{We have demonstrated that spatially resolved spectroscopy has been crucial to confirm the relic nature of J1447-0149, with resolved morphology, kinematic, and stellar population measurements allowing us to constrain the early assembly of the central spheroid and disk of J1447-0149.}


   \maketitle

\section{Introduction}
In the scope of the two-phase formation scenario \citep{2010ApJ...725.2312O}, massive (M$_{\star}\ge 10^{11}$M$_{\odot}$) early-type galaxies (ETGs) formation commences with a first intense and rapid phase at high redshift \citep{Zolotov15}, creating a massive, compact, and passive galaxy (`red nugget', \citealt{Damjanov+11}). 
Recent JWST observations have shown that these quiescent, ultra-compact objects were already formed at redshift $z>3$ \citep{2023MNRAS.520.3974C, 2023Natur.619..716C, 2023Natur.616..266L, 2024NatSR..14.3724N, 2024Natur.628..277G}, and with the most distant one found at $z=7.3$ \citep{2025ApJ...983...11W}.
Then, normally, in the second phase these nuggets undergo a dramatic structural and size evolution due to mergers and interactions, resulting in massive ETGs at $z=0$ \citep{2018A&A...619A.137B}.

However, since merging is stochastic, a small fraction of the high-$z$ red nuggets is expected to survive intact until today, without experiencing significant growth. Known as massive relic galaxies \citep{2009ApJ...692L.118T}, these local red nuggets are mainly composed of “in-situ” material, having assembled the great majority of their stellar mass already by $z\approx2$ and thus showing very old stellar populations ($>$10~Gyr). They are, therefore, ideal laboratories to reveal the early formation processes occurring in the early Universe, but with the ease of being nearby. 

Systematic surveys of ultra-compact massive galaxies (UCMGs, with sizes R$_{\rm e} \le 2$ kpc and masses  $\log(M_{\star}/M_{\odot})\gtrsim$ 10.5) as the ideal relic candidates, have shown that relics are incredibly rare at $z=0$ but their number density rises by almost two orders of magnitude up to $z\sim0.5$ \citep{Trujillo+12_compacts,Poggianti+13_evol, Damjanov+13_compacts, Damjanov+14_compacts, Damjanov+15_compacts,Tortora+16_compacts_KiDS, 2018MNRAS.473..969T, Charbonnier+17_compact_galaxies,Lisiecki_2023}, implying that relics should 
be more common at intermediate redshift. 
Indeed the \textsc{INSPIRE} and E-\textsc{INSPIRE} surveys \citep{DR3, 2025MNRAS.541.2440M} have been able to build a statistically large sample of hundreds of spectroscopically confirmed  UCMGs at $0.05<z<0.5$, and characterise their integrated stellar population properties, based on star-formation histories derived from integrated and seeing-dominated (optical) spectra. 

An interesting result has emerged: the existence of a \textit{degree of relicness (DoR)} among all UCMGs \citep{2017MNRAS.467.1929F, Spiniello21_INSPIRE_I, DR3}. Galaxies with the highest DoR have the most extreme, pristine properties, for example the smallest sizes, shortest star formation histories (SFHs), and the most massive supermassive black holes \citep[SMBHs,][]{2017MNRAS.467.1929F}. Since this definition is based on local ($z=0$) relics, it cannot be easily generalised to more distant UCMGs at higher redshifts for which morphologies and SMBH properties are still elusive. Thus, in this paper, we use the quantitative definition introduced in \citet{DR3}, where the DoR is a dimensionless number ranging from 0 and 1, both for relics and for non-relics, based on their star formation history (SFH). 
The DoR is defined as 
\begin{equation}
    \text{DoR} = \left[f_{M^{\star}_{t\text{BB}=3}}+\frac{0.5~\text{Gyr}}{t_{75}}+\frac{0.7~\text{Gyr}+(t_{\rm Uni}-t_{\rm fin})}{t_{\rm Uni}}\right]\times\frac{1}{3},
    \label{eqn:dor}
\end{equation}
where $f_{M^{\star}_{t\text{BB}=3}}$ is the fraction of stellar mass formed by $z=2$, $t_{75}$ is the cosmic time at which 75 per cent of the stellar mass was in place, $t_{\rm fin}$ is the final assembly time (100 per cent of the stellar mass in place), and $t_{\rm Uni}$ is the age of the Universe at the redshift of the objects\footnote{The values 0.5~Gyr and 0.7~Gyr were chosen based on recent JWST results and such that the DoR ranges between 0 and 1, as explained in \cite{DR3}.}. 

A higher DoR indicates an earlier formation epoch with almost no contribution from younger stars brought in through accretion or formed in later SF episodes;  a lower DoR instead means that, although a fraction of stars are old and were formed during the first phase of the mass assembly, there is a non-negligible percentage ($>25$\%) of later-formed populations with different ages and metallicities.
However, a surprising finding is that, from ground-based multi-band photometry, these galaxies look all very similar despite having a wide range in DoR: they are compact, red and, in the majority of the cases, disk-like \citep{2026arXiv260525075S}. 

A quantity that appears to correlate with the DoR is the stellar velocity dispersion \citep{2023A&A...672A..17D, 2023MNRAS.526.4024G, DR3}. Galaxies with higher DoR tend to exhibit larger velocity dispersions, possibly suggesting a more centrally concentrated mass distribution. However, current measurements are based on integrated, seeing-dominated apertures, which do not allow us to disentangle the true dynamical nature of these systems. In compact galaxies, an elevated integrated velocity dispersion can arise either from a genuinely dispersion-dominated spheroidal structure, or from unresolved rotational support within a fast-rotating disk. 

Spatially resolved measurements of the stellar kinematics are, in general, essential for understanding galaxy formation pathways, as the two-dimensional distribution of stellar velocities and velocity dispersions provides direct constraints on galaxy assembly histories, internal dynamical structure, and the relative importance of dissipative and dissipationless processes \citep[see review in][]{Cappellari2016}. In the specific case of UCMGs, distinguishing between pressure-supported and rotation-supported systems is therefore fundamental for constraining their formation pathways. A spheroidal, pressure-supported system would point toward a rapid and violent assembly, where dissipative processes such as cold gas inflows and gas-rich mergers drive strong central compaction events in the early Universe \citep{2014MNRAS.438.1870D}. In this picture, intense star formation occurs on a short timescale, building a dense stellar core with little subsequent structural evolution. Conversely, a rotationally supported, disk-like system would instead favour a more gradual formation scenario, in which the gas accretes smoothly, settles into a disk, and forms stars in a relatively ordered configuration \citep{2011ApJ...730...38V}.

Hence, resolving the internal kinematics of UCMGs is essential to establish whether the DoR is directly linked to their dynamical state and, ultimately, to their assembly mechanism. Only spatially resolved spectroscopy can break the degeneracy between rotation and pressure support, allowing us to determine whether high-DoR systems are truly the dynamically frozen remnants of early violent compaction events, or whether some of them preserve disk-like structures indicative of a different, but still bursty, evolutionary pathway.

Given the incredibly compact apparent sizes of UCMGs on the sky, spatially-resolved studies have been so far limited to the local Universe, where only three relics have been fully confirmed based on their morphology and stellar population properties \citep{2017MNRAS.467.1929F}. 
\citet{2025MNRAS.540.2555T} carried out the first adaptive-optics (AO) aided observations of a relic galaxy beyond the local Universe, targetting the galaxy KiDS J0842+0059 with SOUL and LUCI at the Large Binocular Telescope. They confirm its compact morphology and disky structure previously inferred from seeing-limited observations \citep{2018MNRAS.473..969T}. 
We now take a further step in the AO-aided study of intermediate-redshift relic galaxies, presenting the first spatially resolved spectroscopic study of a relic galaxy beyond the local Universe capitalising on the AO-aided narrow-field mode (NFM) of the Multi Unit Spectroscopic Explorer \citep[MUSE,][]{2010SPIE.7735E..08B}. 
Throughout this paper, we use cosmological parameters from the \citet{2020A&A...641A...6P} and errors denote the $1\sigma$ uncertainty interval unless stated otherwise.

\section{Data and observations}
\label{sec:data}

\subsection{Target selection}
Our target, the relic galaxy KiDS J1447-0149 (hereafter simply J1447-0149)  at redshift $z=0.2074$, was selected from the final data release of the \textsc{INSPIRE} survey \citep{DR3}, which is based on X-shooter long-slit spectroscopy. 
The selection was driven by both scientific and observational requirements. First, the galaxy had to be observable during ESO Period 115. Second, it had to be sufficiently bright to act as a natural guide star for the adaptive-optics correction, requiring $J_{\mathrm{Vega}}\leq 17$. 
Finally, we restricted the selection to systems with an intermediate DoR, $0.3 < \mathrm{DoR} \leq 0.7$.

This last criterion was chosen to maximise the likelihood of detecting spatial variations in both kinematics and stellar populations. The most extreme relics, with $\mathrm{DoR}>0.7$, are expected to have formed the great majority of their stellar mass rapidly at high redshift and may therefore display only weak signatures of accreted stellar populations. By contrast, an intermediate-DoR object should still retain a dominant old stellar component, while allowing for a measurable contribution from later assembly. J1447-0149 satisfies these requirements: with an integrated DoR of 0.38, it formed $82 \pm 7$ per cent of its stellar mass by $z=2$, comparable to more extreme relics, but required a more extended period, between 6 and 10 Gyr, to complete its assembly \citep{DR3}. This combination of an early dominant formation phase and a more extended residual assembly history makes J1447-0149 an ideal target for a first spatially resolved investigation of relic-galaxy structure beyond the local Universe.

From the seeing-limited X-shooter spectrum, J1447-0149 has an integrated stellar velocity dispersion of $\sigma_{\star,\mathrm{INSPIRE}} = 187 \pm 9\,\mathrm{km\,s^{-1}}$. Its integrated stellar population is old and metal rich, with a regularised mass-weighted age of $9.8\pm0.7\,\mathrm{Gyr}$, $[\mathrm{M}/\mathrm{H}]=0.18\pm0.04$, and a moderate $\alpha$-enhancement, with $[\mathrm{Mg}/\mathrm{Fe}]=0.1 \pm 0.1$ \citep{DR3}. In the KiDS imaging, the galaxy is compact and flattened, with a circularised effective radius of $R_{\rm e}=0.44''$, corresponding to $1.51\,\mathrm{kpc}$, a Sérsic index of $n=3.06$, an axis ratio of $q=0.45$, and a stellar mass of $M_\star=8.66\times10^{10}\,M_\odot$ \citep{2020ApJ...893....4S, DR3}. \citet{2024MNRAS.534.1597S} investigated the local environment of the \textsc{INSPIRE} UCMGs and could only tentatively assign J1447-0149 to a cluster environment.

\subsection{Observations and data reduction}
J1447-0149 was observed with the MUSE at the VLT-UT4 Yepun in conjunction with \textsc{Galacsi} at the Adaptive Optics Facility \citep{2012SPIE.8447E..37S, 2008SPIE.7015E..24A} in NFM as ESO programme P115.27WV (PI J. Hartke). The NFM has a spatial pixel scale of 0\farcs025 and covers a field of view of $7\farcs5 \times 7\farcs5$. Its spectral resolving power varies from 1740 at 4800 \AA\ to 3450 at 9300 \AA. The wavelength range from 5780 \AA\ to 6050 \AA\ is obscured by the Na notch filter to exclude emission from the Na laser guide stars. 

\begin{figure}
    \centering
    \includegraphics[width=8.8cm]{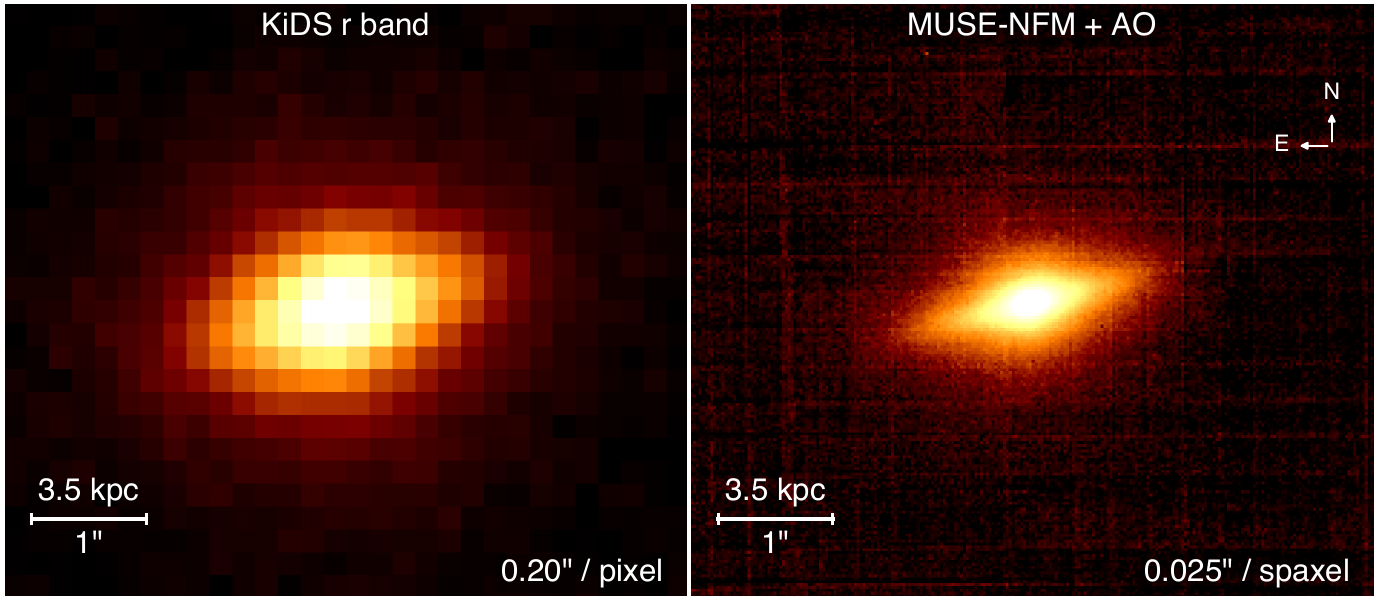}
    \caption{KiDS (\textit{left}) and MUSE-NFM (\textit{right}) $r$-band images covering approximately the same FoV. Note that KiDS has a pixel scale of $0.2\arcsec\,\mathrm{pix}^{-1}$, while the MUSE-NFM  scale is $0.025\arcsec\,\mathrm{spaxel}^{-1}$.}
    \label{fig:kids-vs-nfm}
\end{figure}

Three exposures of 22 minutes each were observed in May and June of 2025. The data were reduced with the \texttt{pymusepipe} wrapper \citep{emsellemPHANGSMUSESurveyProbing2022} of the MUSE data reduction pipeline \citep{weilbacher_2020A&A...641A..28W}. Due to the lack of multiple point sources to align the exposures based on their white-light images, we used the \texttt{image registration}\footnote{\url{https://image-registration.readthedocs.io/}} tool that uses cross-correlation methods \citep{2008OptL...33..156G}. The aligned exposures were then combined into a datacube with the standard tools from the MUSE data reduction pipeline. An $r$-band image obtained from the final data cube is shown in Fig.~\ref{fig:kids-vs-nfm} in comparison to the KiDS image data in the same band, highlighting the impressive improvement in spatial resolution thanks to the AO system and smaller pixel scale. 

\begin{figure*}
    \includegraphics[width=6cm]{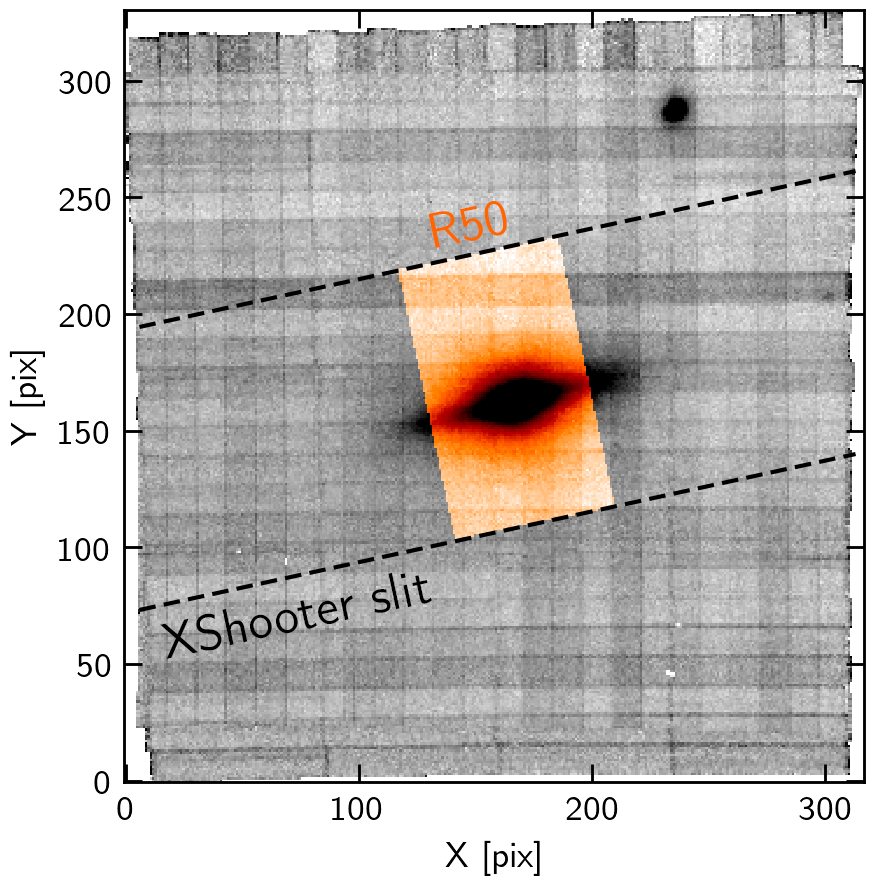}
    \includegraphics[width=12cm]{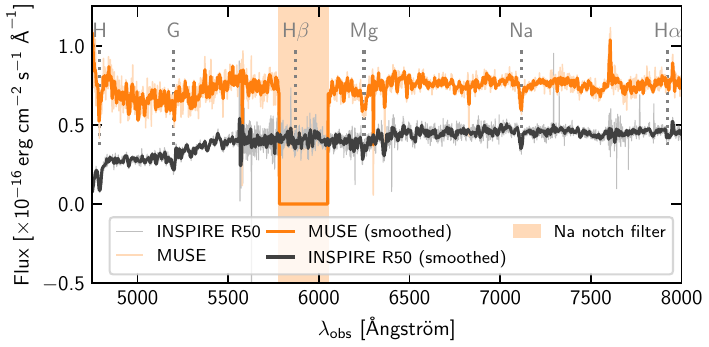}
    \caption{\textit{Left:} White-light image of J1447-0149 with the outline of the X-shooter slit dashed in black, and the location for the spectral extraction done for comparison purposes from both data sets highlighted in orange.  \textit{Right:} The observed spectra, in that same aperture, from \textsc{INSPIRE} DR3 \citep[][gray]{DR3} and extracted from the MUSE datacube in the same aperture (orange), both shown unbinned (thin lines), and median-smoothed to a spectral resolution of 2.5~\AA\ (thick lines), which is the spectral resolution used for the analysis in \citet{DR3}. The labelled dotted vertical lines denote the location of common spectral lines at the redshift of J1447-0149. The shaded orange region denotes the wavelength range obscured by the Na notch filter for the MUSE observations.}
    \label{fig:speccomp}
\end{figure*}

\subsection{Data quality}
To assess the data quality, we extracted spectra in the same aperture as was done in the \textsc{INSPIRE} survey and compare them to the observed spectra from \textsc{INSPIRE} DR3 \citep{DR3} as shown in Fig.~\ref{fig:speccomp}. While the \textsc{INSPIRE} spectrum was observed in seeing-limited conditions, the AO-aided MUSE observations contain a larger fractional contribution from the sky in the same aperture. However, except for a notable sky subtraction residual at $\lambda_\mathrm{obs} \approx 7600$~\AA\ in the spectrum extracted from the MUSE cube, both spectra show the same spectral signatures. We discuss further data quality aspects, such as the image quality as characterised by the point-spread function (PSF), and the reliability of the pipeline errors in Appendix~\ref{app:data}. The adopted PSF has a full width at half maximum (FWHM) of $100\pm3$~mas. 

\section{Analysis}
\label{sec:analysis}
Prior to analysing the spatially-resolved spectroscopic data, we extracted the $r$-band image of J1447-0149 from the datacube, which is shown in the right panel of Fig.~\ref{fig:kids-vs-nfm}. We fit a multi-Gaussian expansion \citep[MGE,][]{2002MNRAS.333..400C} to the stellar surface brightness in this band to determine the photometric centre, position angle, and ellipticity, as well as to better constrain the PSF (see Appendix~\ref{app:data}). Thanks to the higher-resolution photometry compared to previous seeing-limited data, a disk-like photometric component is now visible, motivating us to carry out non-parametric and parametric spheroid-disk decompositions\footnote{Methodologically, this is very similar to the classical bulge-disk decomposition, but we refer to the rounder component as  spheroid, to avoid its interpretation as a classical bulge.}. For the latter, the spheroid is described with a S\'{e}rsic profile \citep{1963BAAA....6...41S}, and the disk with an exponential one, which is a special case of the S\'{e}rsic profile with S\'{e}rsic index $n=1$. 

We carried out the spatially-resolved analysis of our spectroscopic data within the framework of the \texttt{nGIST} pipeline \citep{2025A&A...700A.237F}, based on the Galaxy IFU Spectroscopy Tool \citep[\texttt{GIST},][]{2019A&A...628A.117B}\footnote{\url{http://ascl.net/1907.025}}. We have optimised the \texttt{nGIST} pipeline for the analysis of MUSE-NFM-AO data, by scaling the underestimated variance by a constant when reading the data as detailed in Appendix~\ref{app:data}, and by allowing for a more careful masking of wavelength regions affected by the Na lasers. 

For spatial binning of the data, we used the \texttt{PowerBin} algorithm \citep{2025MNRAS.544.1432C} with different signal-to-noise ratio (S/N) thresholds depending on the aim of the analysis. Prior to binning, we masked all spaxels with a S/N below 1. The \texttt{PowerBin}ned spectra were then log-rebinned with a velocity scale of $41.9\,\mathrm{km}\,\mathrm{s}^{-1}$.
To obtain spatially-resolved kinematics, we used the kinematics module of \texttt{nGIST}, which is a wrapper around the penalised pixel-fitting (\texttt{pPXF}) method \citep{Cappellari04, Cappellari17, 2023MNRAS.526.3273C} with noise estimation, spectral cleaning, and optimal template construction methods as described in \citet{2017ApJ...835..104V}. 

Throughout this analysis, we used template spectra from the MILES single stellar population (SSP) models \citep{2010MNRAS.404.1639V, 2015MNRAS.449.1177V}, spanning a metallicity range from -1.79 to 0.4 dex, and an age range from 0.5~Gyr to the age of the universe at the redshift of J1447-0149 ($\approx 11.2$~Gyr, rounded to 11.5~Gyr), in 0.5~Gyr increments. We chose templates with a bimodal initial mass function (IMF) with a fixed slope of $\Gamma = 1.3$, but acknowledge that the true IMF of relic galaxies such as J1447-0149 may be more bottom-heavy \citep{2015MNRAS.447.1033M,2017MNRAS.467.1929F,2023MNRAS.521.1408M, 2024MNRAS.531.2864M}. The templates are based on BaSTI theoretical isochrones \citep{2004ApJ...612..168P}. We restrict our analysis to a rest-wavelength range of 3976~\AA\ to 6500~\AA, hence avoiding regions with high sky-subtraction residuals and minimising effects due to possible IMF mismatch. 

The derivation of the stellar population properties of the data follows the `\textsc{INSPIRE} philosophy' \citep{Spiniello21_INSPIRE_I, DR3, 2025MNRAS.541.2440M} consisting of a two-step approach, making use of the (\texttt{pPXF}-based) star-formation history  and line-strength modules \citep[based on][]{2006MNRAS.369..497K, 2018MNRAS.475.3700M} of \texttt{nGIST}. 
We first estimated [Mg/Fe] via line-index measurements of the classical magnesium and iron optical features (Mgb, Fe5270 and Fe5335), assuming it can be used as proxy for [$\alpha$/Fe]. We then ran the star-formation history module using only templates with the `correct` $\alpha$-element abundance. This was achieved by linearly interpolating between the [$\alpha$/Fe]$=0$ and [$\alpha$/Fe]$=0.4$ models in steps of 0.1. The only difference with respect to the methodology of \textsc{INSPIRE} is that we used a bootstrapping approach instead of regularisation. This also allows us to obtain uncertainties on our best-fit SFH parameters as detailed in Appendix~\ref{app:errors}.

\section{Results}
\label{sec:results}
In this section, we present the spatially resolved properties of J1447-0149 derived from the MUSE-NFM data. We first describe its morphological properties based on $r$-band imaging obtained from the datacube. We then present its stellar kinematics obtained from the \texttt{PowerBin}ned spectra, focusing on the projected velocity field and the central velocity-dispersion structure. Motivated by these kinematic maps, we define a coarser spatial binning scheme that separates the dispersion-dominated central region from the two sides of the rotating component. This less-granular binning is used to obtain higher-S/N spectra (S/N$\gtrsim40$ per pixel) for the recovery of the higher-order moments of the line-of-sight velocity distribution (LOSVD) and for the subsequent stellar population analysis. Finally, we derive the resolved light-weighted stellar population properties and use the inferred SFHs to estimate the DoR in the centre of the galaxy and in two surrounding regions. 

\subsection{Morphology}
\label{ssec:morph}

\begin{figure*}
    \centering
    \includegraphics[width=18cm]{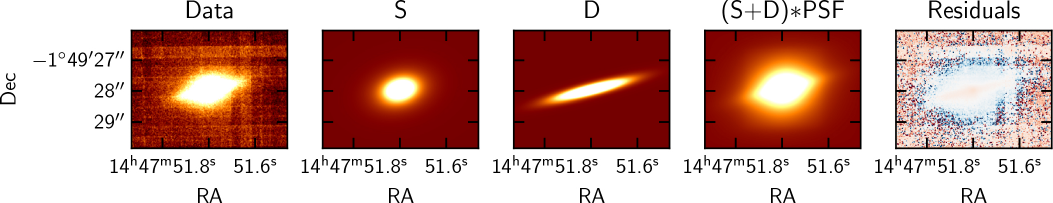}
    \caption{From \textit{left} to \textit{right}: $r$-band image of J1447-0149, best-fit spheroid and disk models, PSF-convolved best-fit model, and fit residuals.}
    \label{fig:2d-SB}
\end{figure*}

We first determined the photometric centre, global position angle $PA = 103.9\degr$, and ellipticity $\epsilon = 0.503$ with \texttt{mgefit} \citep{2002MNRAS.333..400C}, and carried out a non-parametric spheroid-disk decomposition resulting in a spheroid-to-total ratio of $S/T=0.62$, with the spheroid having ellipticities of $\epsilon_{\rm S} = 0.54$ and $\epsilon_{\rm D} = 0.13$ respectively. Based on these measurements, we then used the spheroid-disk parametrisation described in Section~\ref{sec:analysis} to fit two-dimensional surface-brightness profiles. The fit results are shown in Fig.~\ref{fig:2d-SB} and correspond to a spheroid-to-total ratio of $S/T=0.59$, in good agreement with the ratio from the non-parametric model. The spheroid is best described with a S\'{e}rsic index $n_{S} = 2.5$, major-axis effective radius $R_{\mathrm{e},S} = 0\farcs33 \approx 1.17\,\mathrm{kpc}$ and ellipticity $\epsilon_{\rm S} = 0.66$, and the disk with an exponential profile with a scale height of $h_{D} = 0\farcs42 \approx 1.74\,\mathrm{kpc}$ and ellipticity $\epsilon_{\rm D} = 0.14$. These measurements confirm the ultra-compact nature of J1447-0149 that was inferred from seeing-limited data \citep{2024MNRAS.534.1597S}.

The spheroid-to-total ratio is slightly elevated compared to the bulge-to-total ratio local relic sample analysed in \citet{2026arXiv260525075S} that is $B/T = 0.37$. However, owing to deeper photometric data, \citet{2026arXiv260525075S} included more than two structural components in their structural analysis. 
\subsection{Kinematics}
\label{ssec:kin}
\begin{figure}
    \centering
    \includegraphics[width=8.8cm]{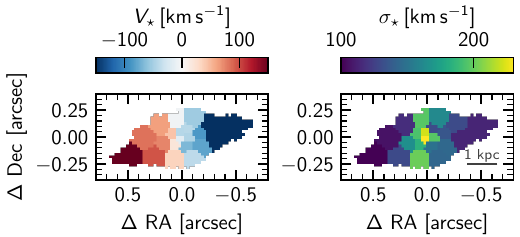}
    \caption{Best-fit kinematic maps for the first two moments of the LOSVD, obtained with \texttt{pPXF} via the kinematics module of \texttt{nGIST} for \texttt{PowerBin}ned data with S/N=10.}
    \label{fig:kinematics}
\end{figure}

\begin{figure*}
    \centering
    \includegraphics[width=18cm]{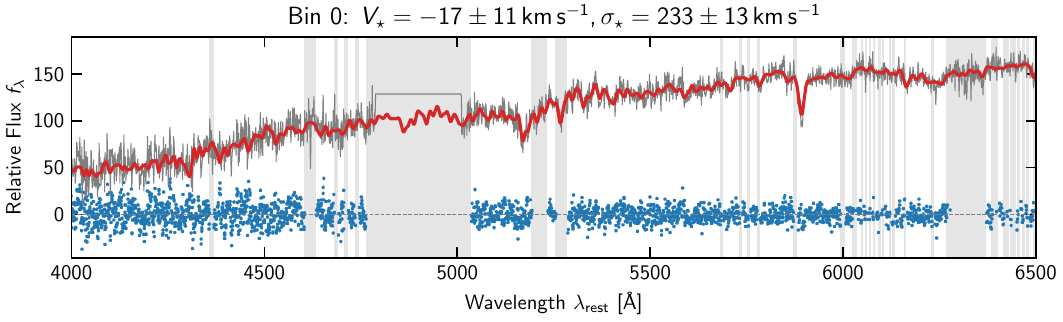}
    \caption{Spectral fit (red) to the data (gray), with residuals shown in blue, for the central bin of the \texttt{pPXF}-based kinematic maps shown in Fig.~\ref{fig:kinematics}. The grey regions denote wavelength intervals that have been masked in the fit.}
    \label{fig:spectral-fit}
\end{figure*}

Figure~\ref{fig:kinematics} shows the stellar velocity and velocity-dispersion fields obtained from the \texttt{PowerBin}ned spectra with $\mathrm{S/N}\geq 10$ per pixel. An example \texttt{pPXF} fit to the central bin is shown in Fig.~\ref{fig:spectral-fit}. The velocity map reveals a clear and spatially coherent gradient across J1447-0149. This pattern is consistent with ordered rotation in the stellar body and demonstrates that the galaxy is not purely pressure supported on the spatial scales probed by the MUSE-NFM data.

At the same time, the velocity-dispersion map shows a pronounced central enhancement. The maximum value, $\sigma_\star = 233\pm13\,\mathrm{km}\,\mathrm{s}^{-1}$, is reached in the innermost region of the galaxy. J1447-0149 therefore combines a rotating stellar component with a dynamically hot central structure. The bootstrapping-based uncertainties on the first two moments of the LOSVD are shown in Fig.~\ref{fig:ppxf-kin-errors} in Appendix~\ref{app:errors}. The amplitude of the velocity gradient and the central peak in $\sigma_\star$ are both larger than the typical uncertainties, indicating that these features are robust within the present analysis.

The central velocity dispersion measured from the MUSE-NFM data is substantially higher than the value obtained from the seeing-limited X-shooter spectrum, $\sigma_{\star,\mathrm{XSH}} = 187 \pm 9\,\mathrm{km\,s^{-1}}$ \citep{DR3}. This difference is not unexpected. The \textsc{INSPIRE} X-shooter measurements are extracted from seeing-dominated spectra and therefore provide luminosity-weighted, aperture-averaged estimates rather than true central values. As discussed in Appendix~A of \citet{Spiniello21_INSPIRE_I} and further tested in \citet{2023A&A...672A..17D}, such integrated velocity dispersions should be regarded as lower limits to the intrinsic central dispersion of compact systems. The MUSE-NFM data therefore reveal that the dynamically hot component of J1447-0149 is more centrally concentrated than could be inferred from the integrated spectrum alone.

\begin{figure*}
    \centering
    \includegraphics[width=18cm]{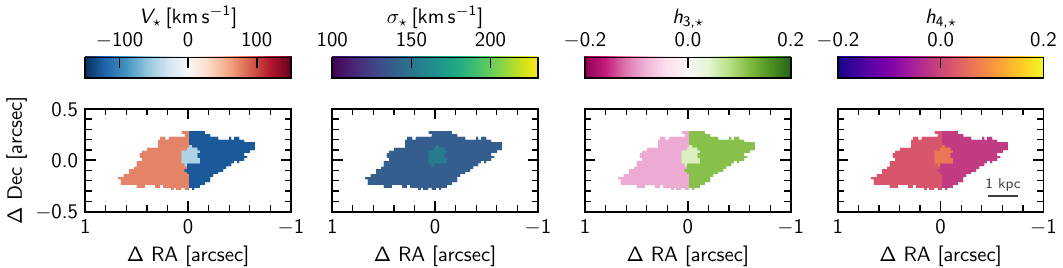}
    \caption{Best-fit kinematic maps for the first four moments of the LOSVD obtained with \texttt{pPXF} via the kinematics module of \texttt{nGIST} for custom-binned data with S/N$\approx40$.}
    \label{fig:kinematics_custom}
\end{figure*}

The kinematic structure, and the need to increase the spectral S/N for detailed stellar populations analysis, motivated a second, coarser binning scheme. In particular, we defined one central bin that covers the dispersion-dominated core, and two outer bins tracing the receding and approaching sides of the rotating component, respectively. This choice sacrifices spatial sampling but increases the $\mathrm{S/N}$ of the extracted spectra, which is essential for a more reliable stellar population modelling and for constraining the higher-order Gauss-Hermite moments of the LOSVD \citep[e.g.][]{2025MNRAS.544.1038F}.

The resulting kinematic maps for the three-bin configuration are shown in  Fig.~\ref{fig:kinematics_custom}. The velocity field retains the same large-scale rotational signature seen in the more granular binning, while the central bin remains the region of highest velocity dispersion. 
The higher-order moments provide additional constraints on the shape of the LOSVD. In particular, $h_3$ changes sign across the kinematic major axis and is anti-correlated with $V_\star$, a behaviour commonly associated with rotating, disc-like stellar components embedded in dynamically hotter systems \citep{1994MNRAS.269..785B, Krajnovic11}. 
The uniformly positive $h_4$ values indicate LOSVDs that are more peaked, and/or have broader wings, than a Gaussian, consistent with either a superposition of cold and hot stellar components or with a degree of radial anisotropy \citep[e.g.][]{1993MNRAS.265..213G, vanderMarel93}.
These higher-order moments should be interpreted with caution, since they are luminosity-weighted averages over relatively large spatial regions and are not by themselves sufficient to determine the full orbital structure. Nevertheless, together with the velocity and velocity dispersion fields, they support the interpretation of J1447-0149 as a compact, high-dispersion core embedded in a rotating stellar structure.

\subsection{Stellar Populations}
\label{ssec:ssp}
\begin{figure*}
    \centering
    \includegraphics[width=18cm]{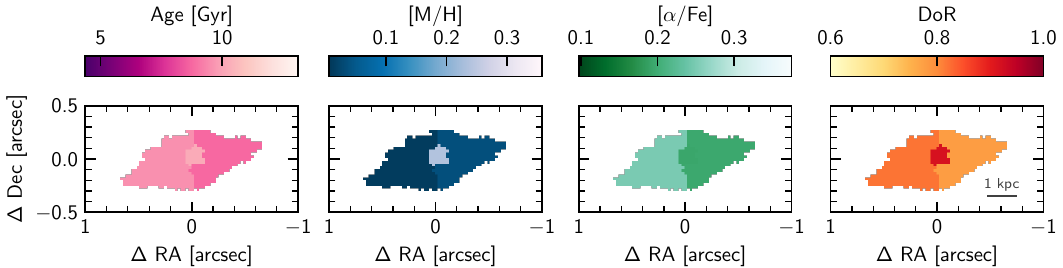}
    \caption{From left to right: light-weighted age and metallicity maps obtained with \texttt{pPXF} via the star-formation history module of \texttt{nGIST}, [$\alpha$/Fe] maps obtained with the line-strength module of \texttt{nGIST}, and the DoR derived from the SFH, all for custom-binned data with S/N$\approx40$.}
    \label{fig:populations+dor}
\end{figure*}

\begin{table*}[]
    \caption{Best-fit kinematics and stellar population properties of J1447-0149 in the three-bin configuration.}
    \centering
    \begin{tabular}{ccccccccc}
    \hline\hline
        Bin & $V_{\star}$ & $\sigma_{\star}$ & $h_{3,\star}$ & $h_{4,\star}$ & Age & [M/H] & [$\alpha$/Fe] & DoR \\
        & [$\mathrm{km\,s^{-1}}$] & [$\mathrm{km\,s^{-1}}$] &  &  & [Gyr] & [dex] & [dex] & \\
        \hline
        Central & ~~$-47 \pm 7$ & $154 \pm 10$ & ~~$0.05 \pm 0.04$ & ~~$0.06 \pm 0.04$ & $10.2 \pm 1.1$ & $0.24 \pm 0.05$ & $0.19 \pm 0.10$ & $0.9^{+0.1}_{-0.2}$ \\
        Right & $-126 \pm 6$ & $139 \pm ~~7$ & ~~$0.11 \pm 0.03$ & $-0.02 \pm 0.04$ & ~~$8.8 \pm 1.0$ & $0.04 \pm 0.08$ & $0.20 \pm 0.10$ & $0.8\pm0.2$\\
        Left & ~~~~$75 \pm 5$ & $139 \pm ~~7$ & $-0.09 \pm 0.03$ & ~~$0.02 \pm 0.04$ &~~$9.6 \pm 0.9$ & $0.01 \pm 0.06$ & $0.24 \pm 0.10$ & $0.8\pm0.2$\\
    \hline\hline
    \end{tabular}

    \label{tab:3bins}
\end{table*}

We derived the stellar population properties of J1447-0149 using the same three spatial bins adopted for the higher-order kinematic analysis. This ensures that the stellar population measurements are matched to the main dynamical components identified above: the central dispersion-dominated region and the two sides of the rotating component. The analysis was performed with the star-formation history and line-strength modules of the \texttt{nGIST} pipeline, following the procedure described in Sect.~\ref{sec:analysis}.

We first determined the $\alpha$-element abundance, finding no variation in the three spatial bins, within the precision allowed by the adopted template grid values  (0.1 dex), with [$\alpha$/Fe]$=0.2$ across the galaxy. 
We then performed the \texttt{pPXF}-based SFH fit after restricting the template library to SSP spectra interpolated to [$\alpha$/Fe]$=0.2$. Finally, we used bootstrapping to obtain uncertainties on the best-fit parameters (see also Appendix~\ref{app:errors}).
The resulting light-weighted age and metallicity maps are shown in the first two panels of Fig.~\ref{fig:populations+dor}. The stellar populations are uniformly old, with only modest spatial variations across the galaxy considering the typical uncertainties of $\approx1\,\mathrm{Gyr}$ (see Fig.~\ref{fig:ppxf-ssp-errors}). The central core is populated by the oldest and most metal-rich stars. The kinematic and stellar-population properties are also summarised in Table~\ref{tab:3bins}.
We repeated the analysis with other SSP libraries (sMILES, \citealp{2023MNRAS.523.3450K} and BPASS, \citealp{2025MNRAS.537.2433B}) and found no systematic differences in the recovered stellar population properties.

To quantify the relic nature of J1447-0149 on resolved scales, we computed the DoR in each bin following Eq.~\eqref{eqn:dor}. Since the DoR is defined in terms of assembled stellar mass as a function of cosmic time, we first converted the \texttt{pPXF} weights from light-weighted to mass-weighted quantities. For each SSP component, the corresponding light fraction was multiplied by the $V$-band mass-to-light ratio of the template with the same age, metallicity, and [$\alpha$/Fe]. 

\begin{figure}
    \centering
    \includegraphics[width=8.8cm]{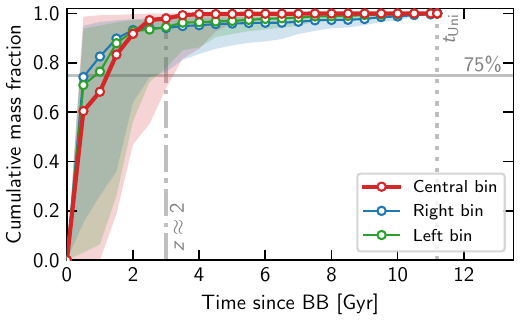}
    \caption{Cumulative mass fractions as a function of time since the Big Bang for the three spatial bins, with the central bin shown in red and the outer bins in green and blue respectively. Circles denote the age grid. The shaded regions cover the  16th–84th percentile range. The gray dash-dotted (dotted) vertical line denotes $z\approx2$ ($t_{\rm Uni}$, i.e. the age of the Universe at the redshift of J1447-0149) and the horizontal line marks 75 per cent of the cumulative mass fraction.}
    \label{fig:SFH-bins}
\end{figure}

The resulting DoR map is shown in the rightmost panel of Fig.~\ref{fig:populations+dor}. The highest DoR, $0.9^{+0.1}_{-0.2}$, is found in the central bin, indicating that the dispersion-dominated core is the most relic-like component of J1447-0149. 
The two outer bins, associated with the rotating component, show slightly lower DoR values ($0.8\pm0.2$ and $0.8\pm0.2$, respectively). Uncertainties on these values were obtained with bootstrapping, and the error maps are shown in Fig.~\ref{fig:ppxf-ssp-errors} in Appendix~\ref{app:errors}. 

The corresponding cumulative mass assembly histories derived from these mass-weighted weights are shown in Fig.~\ref{fig:SFH-bins} for each bin. 
The uncertainties on the SFHs (obtained with bootstrapping) are rather large and indicate that we cannot fully distinguish the SFHs between the central (red line) and outer bins (blue and green lines). 
This is due to the moderate S/N of the data, but also due to the limited age grid resolution (0.5~Gyr) in the early Universe, as indicated by the open circles in Fig.~\ref{fig:SFH-bins}.
However, even when considering these limitations, our analysis indicates that J1447-0149 is a relic of the early Universe, as more than 90 per cent of its mass was already in place two Gyr after the Big Bang, and its core was fully assembled by $z\approx2$. The moderate depth of the data implies that we are only probing the brightest, most central, and highest DoR regions, while the seeing-limited X-shooter data covering a larger area, resulted in a DoR-value of 0.38. With the current data, we cannot rule out the presence of younger and less relic-like stellar populations at large radii.

\section{Discussion}
\label{sec:discussion}

The spatially resolved data reveal that J1447-0149 is not described by a single morphological and dynamical component. Its velocity field shows a coherent gradient across the galaxy together with a compact region of enhanced velocity dispersion. The change in sign of $h_3$ and its anti-correlation with $V_\star$ support the presence of an ordered rotating component superimposed on a hotter stellar distribution \citep{1994MNRAS.269..785B}. The positive $h_4$ values are not unique evidence for two components, because they may also reflect orbital anisotropy \citep{1993MNRAS.265..213G}; nevertheless, together with the lower-order LOSVD moments and galaxy morphology, they are consistent with a dynamically hot centre embedded in a disk-like rotating stellar body.

The stellar populations follow the same qualitative decomposition. The central region contains the nominally oldest and most metal-rich population and reaches the highest resolved DoR, whereas the outer bins remain old, $\alpha$-enhanced, and highly relic-like but have somewhat more extended recovered SFHs. Because the central population bin extends beyond the narrow peak of the velocity-dispersion field, it includes light from the surrounding rotating structure; this mixing can dilute an intrinsic central population contrast. Since the SFH and DoR uncertainties overlap, these two measurements do not establish a statistically significant separation between the formation histories of the central and outer regions. However, given the negative age and metallicity gradients, we can infer that the centre formed slightly earlier with more efficient star formation than the outskirts. The overall flat [$\alpha$/Fe] distribution indicates that centre and outskirts had similar overall formation timescales. The dynamically hot centre thus coincides with the nominally oldest, most metal-rich, and most rapidly assembled stellar population.

A useful local comparison is NGC~1277, the most extreme relic galaxy in the local Universe \citep{2014ApJ...780L..20T}. Like J1447-0149, NGC~1277 is a compact fast rotator with a pronounced central velocity-dispersion peak, while its rotational velocity increases towards larger radii \citep{2014ApJ...780L..20T,2023A&A...675A.143C}. Its stellar population is uniformly old, with only mild radial variations in age and $\alpha$-enhancement, although its metallicity decreases outwards and it has assembled the bulk of its stellar mass much quicker \citep{2014ApJ...780L..20T, 2015MNRAS.451.1081M, 2017MNRAS.467.1929F}. The centre of J1447-0149 therefore resembles the local relic prototype in combining ordered rotation with a centrally concentrated high-dispersion component, while its resolved stellar populations further indicate that this dynamically hot centre hosts the nominally oldest and most metal-rich stars. We note, however, that our data only covers $\approx 1.4\,R_{\rm e}$, while the stellar population properties of the three local relic galaxies NGC~1277, Mrk~1216, and PGC~032873 are characterised up to $\approx4\,R_{\rm e}$ effective radii. Within the central $\approx 1-2\,R_{\rm e}$ their stellar population properties are relatively similar \citep{2017MNRAS.467.1929F}.

The main implication is therefore that relics do not 
need to be associated with a purely pressure-supported or dynamically homogeneous system. The coexistence of a dynamically hot centre and ordered rotation suggests that the early dissipative event that built J1447-0149 and NGC~1277 did not completely erase angular momentum. 
From an evolutionary point of view, both wet compaction and gas-rich mergers provide plausible routes to this configuration. In wet-compaction models, dissipative inflows in gas-rich, perturbed discs produce compact central starbursts followed by inside-out quenching, while an extended component may survive or develop subsequently \citep{2014MNRAS.438.1870D,Zolotov15}. Gas-rich merger simulations likewise produce massive, kiloparsec-scale quiescent remnants with high velocity dispersions and substantial rotation \citep{2010ApJ...722.1666W}. 
The present observations do not distinguish among these mechanisms, but require a formation pathway capable of producing a dense, dynamically hot centre without removing all of the angular momentum of the surrounding stellar structure.

Observations at high redshift show that the required ingredients can coexist before and after quenching. JWST rest-frame near-infrared imaging indicates that massive star-forming galaxies at cosmic noon can already host prominent central stellar structures \citep[e.g.][]{2024ApJ...974L..28B, 2025NatAs...9..141B}. In the quiescent phase, several lensed compact massive galaxies at $z\approx2$ show significant rotation \citep{2015ApJ...813L...7N, toft_massive_2017, 2018ApJ...862..126N}, demonstrating that substantial ordered rotation can persist after quenching. J1447-0149 may therefore be the descendant of a system in which a dense central component and significant net angular momentum were both established at high redshift, akin to the fast rotators at cosmic noon \citep[e.g.][]{2025A&A...702A.110S} and beyond \citep{2026MNRAS.547ag210P} whose kinematics have been characterised with JWST. 

Its subsequent evolution must, however, have remained limited. Following compact systems from $z=2$ to the present day, \citet{2016MNRAS.456.1030W} found that some become the compact cores of larger galaxies through the acquisition of ex-situ envelopes, whereas others remain comparatively undisturbed. J1447-0149 appears more closely related to the latter population: its compactness, predominantly old stellar populations, and preserved rotation argue against substantial late mass and size growth. This inference is also consistent with simulations in which rotational support decreases as the local ex-situ fraction increases and major mergers suppress the rotation of the in-situ component \citep{2021A&A...647A..95P}. Controlled merger sequences similarly show that repeated mergers can transform initially fast-rotating compact progenitors into slow rotators, although the outcome depends on the initial structure and merger history \citep{2024MNRAS.535.1202R}. The coherent rotation of J1447-0149 therefore disfavours an extensive sequence of dry mergers.

A limited amount of later evolution remains possible. One or a small number of minor accretion events could have added stars outside the compact centre without destroying the galaxy's compactness or erasing its ordered rotation. Alternatively, the somewhat more extended SFHs of the outer bins could reflect residual in-situ star formation after the main central formation event. The present data cannot distinguish between these possibilities, because the DoR constrains when the observed stars formed, but not whether they formed in situ or were subsequently accreted. The slightly lower outer DoR should therefore not be interpreted by itself as evidence for an ex-situ envelope. At the same time, the uniform $[\alpha/{\rm Fe}]$ and early cumulative SFHs show that the rotating structure is not a young disc added at late times: both the central and outer regions formed the majority of their stellar mass early.

J1447-0149 thus appears to preserve a compact, exceptionally old, dynamically hot centre together with an old stellar structure that retains net angular momentum. The contrast between its central and aperture-averaged velocity dispersions further shows that integrated spectroscopy can dilute centrally concentrated dynamical signatures. Spatially resolved observations of a larger sample will be required to determine whether the correlation between DoR and integrated $\sigma_\star$ \citep{2023A&A...672A..17D} is systematically associated with compact high-dispersion centres, reduced rotational support, or, a combination of both.

\section{Summary and Conclusions}
\label{sec:conclusions}

This paper presents the first spatially resolved spectroscopic study of a relic galaxy beyond the local Universe. 

We targeted J1447-0149 at $z=0.2074$ with AO-assisted MUSE-NFM observations. J1447-0149 was selected from the final data release of the \textsc{INSPIRE} survey as an intermediate-DoR system, with an early dominant formation phase but a sufficiently extended residual assembly history to make internal kinematic and stellar-population variations potentially detectable. 

Our main results are as follows.

\begin{itemize}
    \item The AO-aided MUSE-NFM data spatially resolve the stellar body of this compact galaxy and reveal a clear, coherent velocity gradient. J1447-0149 is therefore not a purely pressure-supported system, but hosts an ordered rotating stellar component surrounding a compact, dynamically hot core. The velocity-dispersion field reaches $\sigma_\star=233\pm13\,\mathrm{km\,s^{-1}}$ in the innermost region, substantially above the seeing-limited X-shooter value ($\sigma_{\star, \rm INSPIRE} =187 \pm 9\,\mathrm{km\,s^{-1}}$), confirming that the seeing-limited integrated spectrum underestimates the central dynamical concentration.

    \item The higher-order LOSVD moments support the composite kinematic picture. In the three-bin configuration, $h_3$ changes sign across the kinematic major axis and is anti-correlated with $V_\star$, as expected for a rotating stellar component embedded in a hotter system. The positive $h_4$ values indicate LOSVDs that are more centrally peaked, and/or have broader wings than a Gaussian. While these quantities are luminosity-weighted averages over relatively large spatial regions, the combined $V_\star$, $\sigma_\star$, $h_3$, and $h_4$ measurements point to a compact high-dispersion core embedded in a rotating stellar structure.

    \item The resolved SFHs show that the dispersion-dominated core is the most relic-like component of J1447-0149. The central bin assembled its stellar mass extremely early, within $\sim 2\,\mathrm{Gyr}$ after the Big Bang, and reaches the highest DoR, $\mathrm{DoR}=0.9^{+0.1}_{-0.2}$. The two outer bins, associated with the rotating stellar structure, also assembled their stellar mass very early, but have more extended assembly histories, resulting in a slightly lower DoR value in both bins, $\mathrm{DoR}=0.8\pm0.2$. 

\end{itemize}
We have shown that spatially resolved integral-field spectroscopy is essential for uncovering the internal dynamics, morphology, and stellar population content of relic galaxies. While the data of J1447-0149 are relatively shallow, not allowing for a fully spatially resolved stellar population analysis, or detailed dynamical and mass models, we constrained the early assembly of its most central components. Our analysis is a stepping stone for future studies of intermediate-redshift relic galaxies with instruments such as MUSE-NFM.

\begin{acknowledgements}
We thank the staff at ESO Paranal for carrying out our observations and Giacomo Beccari from the User Support Department for his support of our programme. J.H., C.S., E.C., and A.F.M. acknowledge the financial support from the visitor and mobility program of the Finnish Centre for Astronomy with ESO (FINCA) as well as funding to support the scientific use of our ESO data from FINCA. J.H acknowledges CSC -- IT Center for Science, Finland, for computational resources, support from TCSMT in the form of a starting grant. Part of this work was carried out during a visit to ESO Chile supported by the ESO Scientific Visitor Programme. 
J.H. thanks Amelia Fraser-McKelvie and Jesse van de Sande for useful discussions about the use of the \texttt{nGIST} pipeline. 
J.H. thanks her former colleagues from the IRLOS+ upgrade team that made observing relic galaxies with MUSE-NFM possible.
A.F.M. acknowledges support from RYC2021-031099-I and PID2024-162088NB-I00 of MICIN/AEI
G.D. acknowledges support by UKRI-STFC grants: ST/T003081/1 and ST/X001857/1.
This research is partially based on data from the MILES project.
Based on observations collected at the European Southern Observatory under ESO programme P115.27WV. Artificial Intelligence (OpenAI gpt-5.2 model, accessed via the UTU Staff Chat tool) was consulted for streamlining \texttt{python} codes for data visualisation. 
In addition to the software mentioned in the text, this research made use of \texttt{astropy} \citep{2013A&A...558A..33A,2018AJ....156..123A}, \texttt{astroplan} \citep{2018AJ....155..128M}, \texttt{astroquery} \citep{2019AJ....157...98G}, \texttt{esorex} \citep{2015ascl.soft04003E}, \texttt{matplotlib} \citep{Hunter:2007}, \texttt{mpdaf} \citep{2019ASPC..521..545P}, \texttt{numpy} \citep{2011arXiv1102.1523V}, and \texttt{photutils} \citep{larry_bradley_2023_7946442}.
\end{acknowledgements}

%

\bibliographystyle{aa}
\bibliography{relics.bib}

\begin{appendix}




\onecolumn
\FloatBarrier
\section{Data quality}
\label{app:data}
\subsection{Image quality and PSF}
The only point-like source in the image is a foreground star in the top-right corner of the data cube, however, since it is well outside of the radius in which we expect a good correction by the AO system, we do not use it to estimate the PSF. 
Instead, we may be able to use the nuclear [\ion{N}{ii}] emission to constrain the PSF in the centre of the FoV, under the assumption that the nuclear emission is point-like. We extract a pseudo-narrow-band image (14 Å wide) around the [\ion{N}{ii}] emission at rest-frame wavelength 6583 Å, subtracting two narrow continuum slices on each side of the emission line, which is shown in the leftmost panel of Fig.~\ref{fig:psf-comp}. 

\subsubsection{PSF fitting with \texttt{maoppy} and comparison with single Moffat estimates}
\begin{figure*}
    \centering
    \includegraphics[width=18cm]{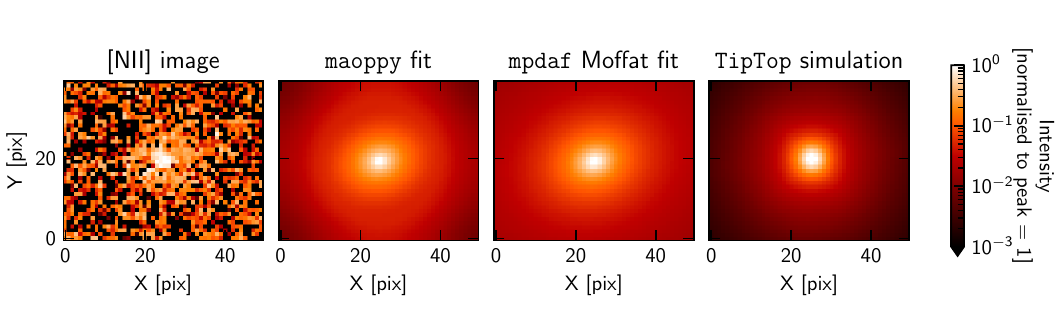}
    \caption{Comparison of the continuum-subtracted [NII] narrow-band image (\textit{leftmost panel}) with PSF estimates at the same wavelength obtained with, from \textit{left to right}, \texttt{maoppy} (top), \texttt{mpdaf}, and \texttt{Tiptop}. }
    \label{fig:psf-comp}
\end{figure*}

\begin{figure}
    \centering
    \includegraphics[width=8.8cm]{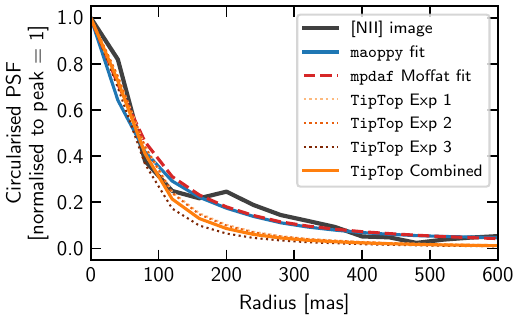}
    \caption{Circularised PSF estimates from \texttt{maoppy} (blue line), \texttt{mpdaf} (red dashed  line), and \texttt{TipTop} (dotted orange lines for individual exposures and solid orange line for the combined estimate) in comparison to a cutout of the continuum-subtracted [\ion{N}{ii}] image, whose radial profile is denoted by the black line.}
    \label{fig:psf-comp-1d}
\end{figure}

For MUSE-NFM, physically motivated PSFs can be fit with \texttt{maoppy} \citep{2019A&A...628A..99F}, but likely due to the very shallow data not revealing the expected turbulent halo of the PSF, the fitting routine could not constrain the Fried parameter $r_0$, even when fixing the frequency transition to $\alpha=5\times10^{-2}\,\mathrm{m}^{-1}$ as suggested in the documentation. We fitted the PSF with a fixed $r_0$, with the resulting profile indistinguishable from a single 2D Moffat, which we fitted to the data with \texttt{mpdaf}, as illustrated in Figs.~\ref{fig:psf-comp} and \ref{fig:psf-comp-1d}. The FWHM of the fitted 2D Moffat profile are (0\farcs12, 0\farcs095) and of the \texttt{maoppy} fit are (0.\farcs1, 0\farcs. For comparison, the effective radius of the galaxy is $0\farcs44$. 

\subsubsection{PSF prediction with TipTop}
We have also predicted the MUSE-NFM PSF with the \texttt{TipTop} tool, which has been developed to predict the PSF of ELT instruments \citep{2024SPIE13097E..2YN}. To derive a realistic PSF model for each of the three exposures, we used the time-averaged $C_n^2$ and wind profiles that are readily available from the raw data headers, and calibrated the additional aberrations as described in \citet{2026arXiv260804819C}. The resulting predictions for each exposure are shown with red and orange lines in Fig.~\ref{fig:psf-comp-1d}, in comparison to the best-fit PSF obtained with \texttt{maoppy} shown in blue. The combined PSF from the three exposures is shown in the rightmost panel of Fig.~\ref{fig:psf-comp}.

\subsection{Adopted PSF}
\begin{table}[]
    \centering
    \begin{tabular}{lll}
        \hline 
        \hline 
        Model & circularised FWHM & Strehl ratio \\
         & [mas] & [per cent] \\
        \hline 
        \texttt{maoppy} & $127.8\pm0.6$ & 0.8\\
        Single Moffat & $153\pm10$ & -- \\
        \hline 
        \texttt{Tiptop}-Exp 1 & 104 & 5.7 \\
        \texttt{Tiptop}-Exp 2 & 104 & 5.8 \\
        \texttt{Tiptop}-Exp 3 & ~~93 & 8.2 \\
        \texttt{Tiptop}-combined & $100\pm3$  & $6.6\pm0.7$ \\
        \hline 
        \hline 
    \end{tabular}
    \caption{Parameters of the PSF estimates obtained with different methods, the horizontal line distinguishing values obtained from image fitting and from AO telemetry.}
    \label{tab:IQ}
\end{table}

\begin{figure*}
    \centering
    \includegraphics[width=18cm]{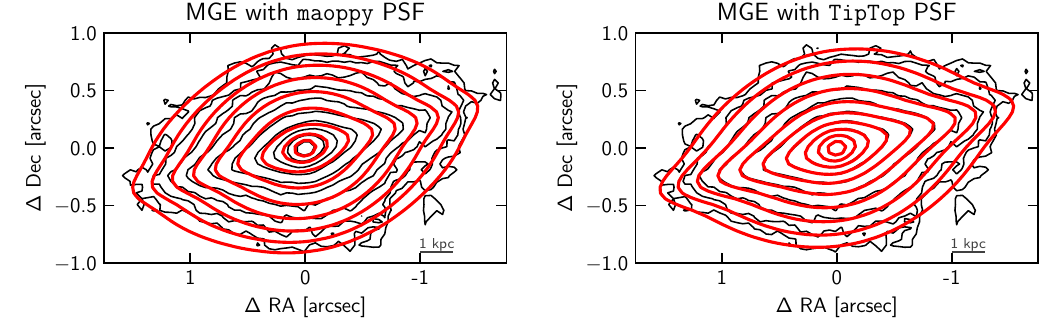}
    \caption{MGE fits (red contours) to the $r$-band image of J1447-0149 (black contours), using the \texttt{maoppy}-based PSF (\textit{left}) and the PSF predicted with \texttt{TipTop} (\textit{right}), illustrating that the latter results in a better agreement between the MGE and the data.}
    \label{fig:psf+mge}
\end{figure*}
We report the best-fit parameters of the PSF estimates in Table \ref{tab:IQ}, noting that the different methods do neither agree on the Strehl ratios nor on the FWHM estimates.
Both PSF estimates have issues: on the one hand, we cannot rule out that the strong wings of the \texttt{maoppy}-PSF are due to fitting it to a noisy [NII] image. On the other hand, TipTop is known to overpredict the GALACSI AO performance, resulting in 'too sharp' PSFs \citep{2026A&A...708A.311K}. However, when performing an MGE fit to the $r$-band image, it is clear that the \texttt{maoppy}-based PSF does not result in a good fit to the imaging data. We therefore adapt the \texttt{TipTop}-based PSF, which has a FWHM of $100\pm3$ and corresponds to observations with a Strehl ratio of $6.6\pm0.7$ per cent. 

\subsection{Spectral quality and errors}
\subsubsection{Long spectra}
\begin{figure*}
    \centering
    \includegraphics[width=18cm]{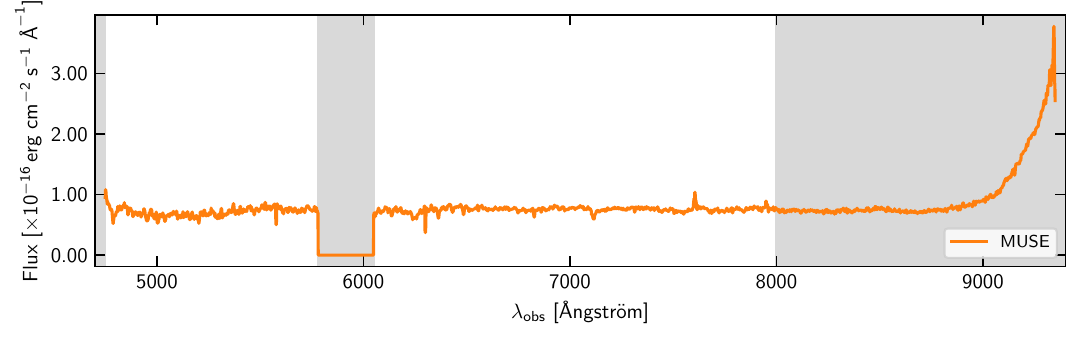}
    \caption{Spectrum extracted within $R_{50}$ in an X-shooter-like aperture as indicated in the left panel of Fig.~\ref{fig:speccomp} extracted over the full MUSE wavelength range. Shaded grey regions are masked prior to the scientific analysis of the data.}
    \label{fig:longspec}
\end{figure*}
Figure~\ref{fig:longspec} shows the spectrum extracted within $R_{50}$ in an X-shooter-like aperture over the full MUSE wavelength range from 4800 to 9300~\AA. We note that this aperture is much larger than the one used for the scientific analysis of the data, and, thanks to the superior spatial resolution compared to the seeing-limited X-shooter data,  also contains some sky background. This allowed us to identify a region redward of 8000~\AA\ where the sky background rapidly increases. We therefore masked this region, as well as a small wavelength region bluer than 4749~\AA.   
Finally, the spectral region covered by the Na notch filter to avoid contamination from the AO laser guide stars was also masked prior to any further analysis. All masked spectral regions are indicated in grey on Fig.~\ref{fig:longspec}. 

\subsubsection{Variance cube}
\begin{figure*}
    \centering
    \includegraphics[width=18cm]{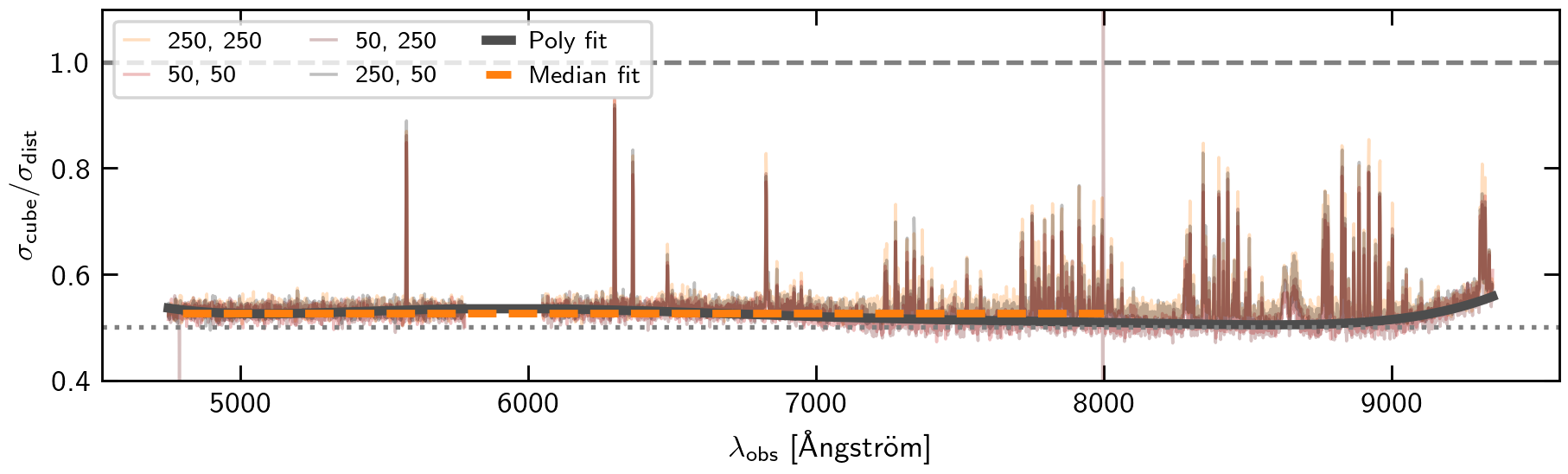}
    \caption{Ratio of the errors as propagated by the MUSE data reduction pipeline ($\sigma_\mathrm{cube}$) and the standard deviation of fluxes in a set of four square patches, whose central coordinates are shown in the legend. The solid gray line indicates a 6th-order polynomial to the ratio and the dashed orange line a median fit for wavelengths $\lambda_\mathrm{obs} \leq 8000\,$Å. The dashed (dotted) horizontal gray line indicates a ratio of 1 (0.5).}
    \label{fig:variance-scaling}
\end{figure*}
Following the discussion in \citet{2025MNRAS.543.1373P}, we calculated the ratio of the pipeline errors (via the variance map, $\sigma_\mathrm{cube}$) and the distribution of fluxes in a set of square patches (20$\times$20 pix, $\sigma_\mathrm{dist}$), also finding that the pipeline-produced variance map underestimates that true variance, as illustrated by the thin lines on Fig.~\ref{fig:variance-scaling}. We first fit a 6th-order polynomial to the ratio after filtering out high residuals at the location of known sky lines, again following \citet{2025MNRAS.543.1373P}, resulting in: 
\begin{equation}
    P(x) = 0.0002 -  0.0096x + 0.1621x^2 -  1.4545x^3 + 7.2849x^4 -  19.2407x^5 + 21.3832x^6, 
\end{equation}
where $x = \frac{\lambda}{1000~\mathrm{Å}}$. However, since the fitted polynomial does not vary significantly over the wavelength range to which we restricted our analysis, we decided to scale the variance cube by a constant value of $0.525^2$. This value corresponds to the square of the median filtered error ratio, which is indicated by the dashed orange horizontal line on Fig.~\ref{fig:variance-scaling}. 

\FloatBarrier
\section{Error maps}
\label{app:errors}
\begin{figure}
    \centering
    \includegraphics[width=8.8cm]{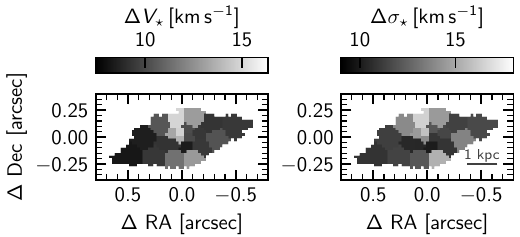}
    \caption{Error maps for fits of the stellar velocity $V_\star$ and velocity dispersion $\sigma_\star$ obtained with bootstrapping for the \texttt{PowerBin}ned spectra with S/N=10 per pixel.}
    \label{fig:ppxf-kin-errors}
\end{figure}

\begin{figure*}
    \centering
    \includegraphics[width=18cm]{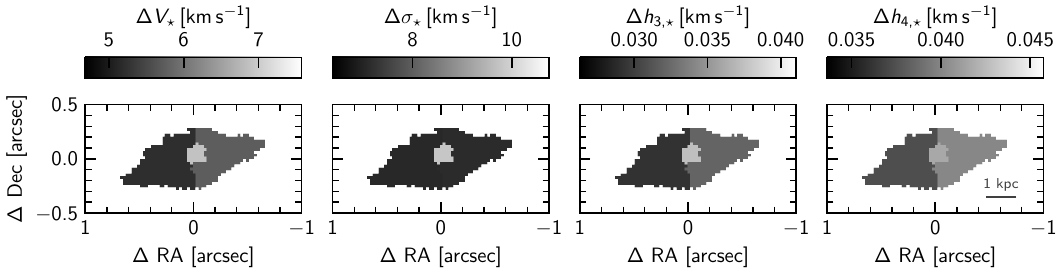}
    \caption{Error maps for fits of the first four moments of the LOSVD obtained with bootstrapping for the \texttt{PowerBin}ned spectra with S/N$\approx40$ per pixel.}
    \label{fig:ppxf-kin-errors-custom-bins}
\end{figure*}

\begin{figure*}
    \centering
    \includegraphics[width=18cm]{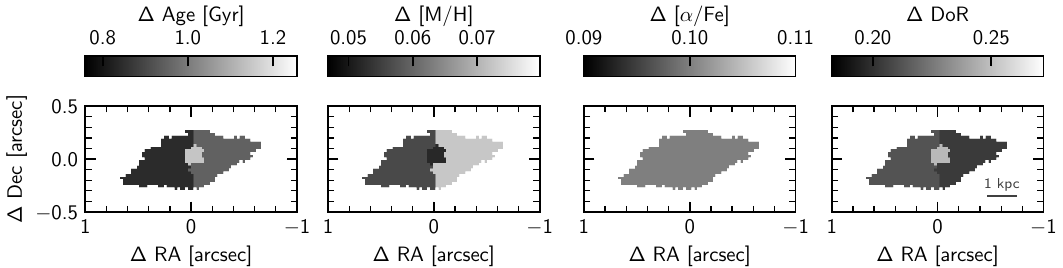}
    \caption{Error maps for fits of the stellar population parameters age and metallicity [M/H] obtained with bootstrapping for the \texttt{PowerBin}ned spectra with S/N$\approx40$ per pixel, a constant error on [$\alpha$/Fe] based on the model grid spacing, and error on the DoR, again from the bootstrapped SFHs.}
    \label{fig:ppxf-ssp-errors}
\end{figure*}

In this appendix, we present the error maps derived with the nGIST pipeline. While we use the architecture for Monte Carlo (MC) realisations as detailed in \citet[][Appendix A]{2025A&A...700A.237F}, instead of perturbing spectra with Gaussian noise with zero mean and a wavelength-dependent standard deviation corresponding to the spectral error \citep{2023A&A...673A.147P}, we used bootstrapping. In particular, we re-ran \texttt{pPXF} 200 times with input spectra that were obtained by `wild' bootstrapping \citep{DAVIDSON2008162} the best-fit spectra obtained from a prior \texttt{pPXF} run with a moderate amount of regularisation (\texttt{regul} = 1) as detailed in \citet{2023MNRAS.526.3273C}. Figure~\ref{fig:ppxf-kin-errors} shows the error maps corresponding to the kinematic maps presented in Fig.~\ref{fig:kinematics} in Sect.~\ref{ssec:kin}, where we fit the first two moments of the LOSVD   and Fig.~\ref{fig:ppxf-kin-errors-custom-bins} shows the error maps corresponding to Fig.~\ref{fig:kinematics_custom}, where we fit the first four moments of the LOSVD to spectra with a higher S/N. Figure~\ref{fig:ppxf-ssp-errors} shows the errors on the age and metallicity maps presented in the two leftmost panels of Fig.~\ref{fig:populations+dor} in Sect.~\ref{ssec:ssp}. 
\label{app:errors}

\end{appendix}
\end{document}